\documentclass[]{jpconf}
\usepackage{graphicx}
\usepackage{feynmp-auto}
\usepackage{amsmath}
\usepackage{float}
\usepackage{hyperref}
\usepackage{subfig}
\usepackage{xspace}
\usepackage{multirow}

\renewcommand{\pt}{\ensuremath{p_\text{T}}\xspace}

\begin{document}
\title{Exploring LHC Run 1 and 2 data using the Madala hypothesis}
\author{Stefan von Buddenbrock}
\address{School of Physics, University of the Witwatersrand, Wits 2050, South Africa}

\ead{stef.von.b@cern.ch}

\begin{abstract}
The Standard Model (SM) Higgs boson, with its experimental discovery in 2012, has long been an interesting particle to study with the intention of exploring new physics ideas beyond the SM (BSM). Its properties are still not well understood, and there are several features in LHC Run 1 and Run 2 data which point at the possibility of extensions to the SM Higgs sector. This work explores the Madala hypothesis, which is the introduction of a heavy scalar (the Madala boson) to the SM, in addition to a real scalar $S$ and dark matter (DM) candidate $\chi$. This hypothesis has previously been used to explain several anomalous features observe in the LHC Run 1 data. This work extends the study to Run 2 data, and shows that the particle spectrum predicted in the Madala hypothesis is indeed compatible with LHC data. Further study prospects and striking signatures for searches are presented.
\end{abstract}

\section{Introduction\label{sec:intro}}

In 2012, the Standard Model (SM) had its particle spectrum completed by the discovery of the SM Higgs boson ($h$) by the ATLAS~\cite{Aad:2012tfa} and CMS~\cite{Chatrchyan:2012xdj} collaborations at the Large Hadron Collider (LHC). The Madala hypothesis is an extension of the SM, and is one of the many hypotheses in the literature which predicts physics beyond the SM (BSM). At its core, the Madala hypothesis extends the SM by introducing two new scalars that are heavier than the SM Higgs boson. As discussed below, the hypothesis is merely a simplified model with the aim of explaining several particular features of the LHC Run 1 and 2 data. It can be discussed in the context of UV-complete and gauge symmetric BSM scenarios that predict extra scalars (such as 2HDMs, lefr-right symmetric models, etc.), but the focus of this short paper is to treat it as simply as possible with the purpose of exploring the data.

\begin{figure}
\centering
\subfloat[]{\includegraphics[width=0.3\textwidth]{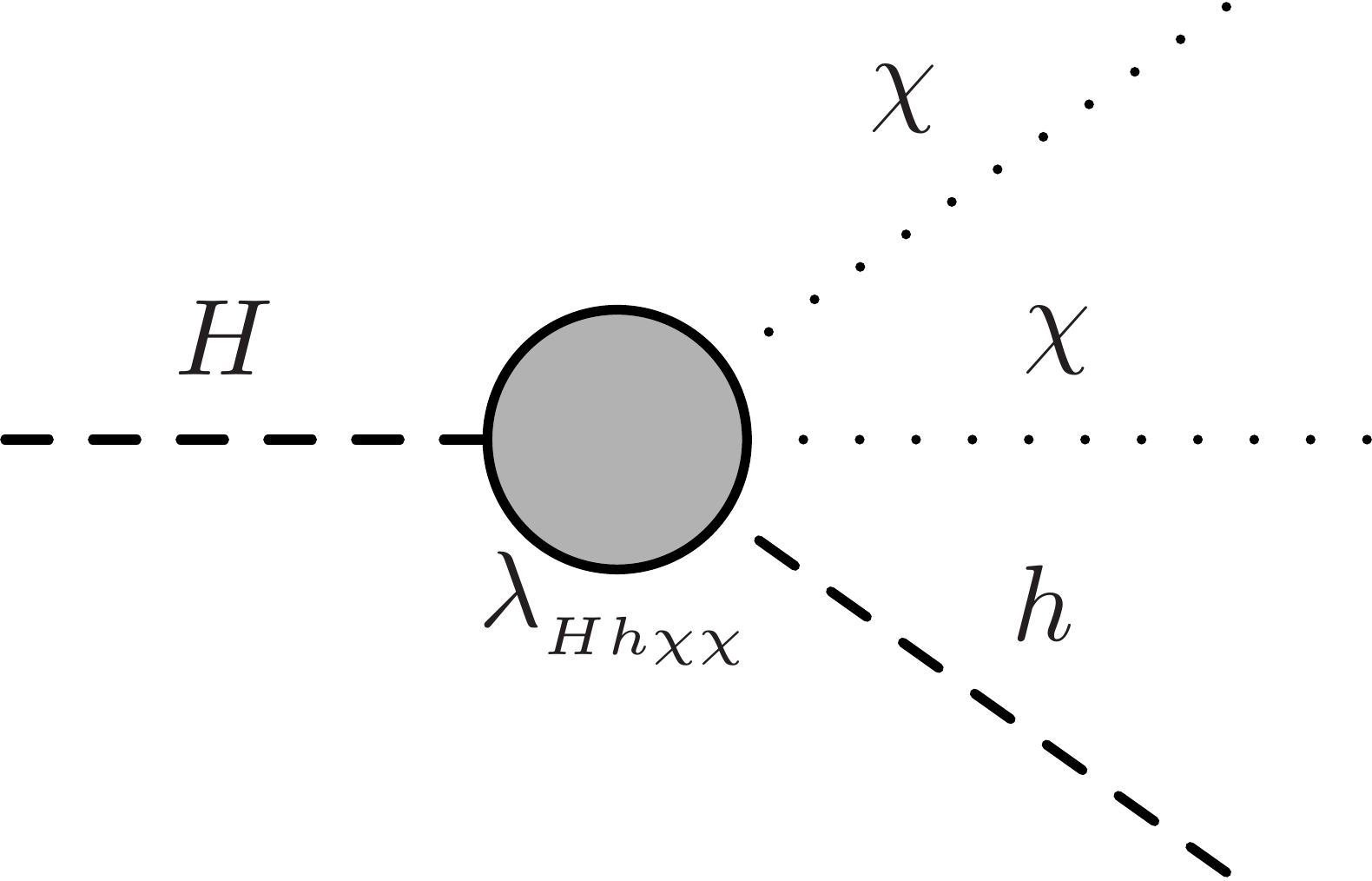}}
\quad\quad\quad\quad\quad
\subfloat[]{\includegraphics[width=0.3\textwidth]{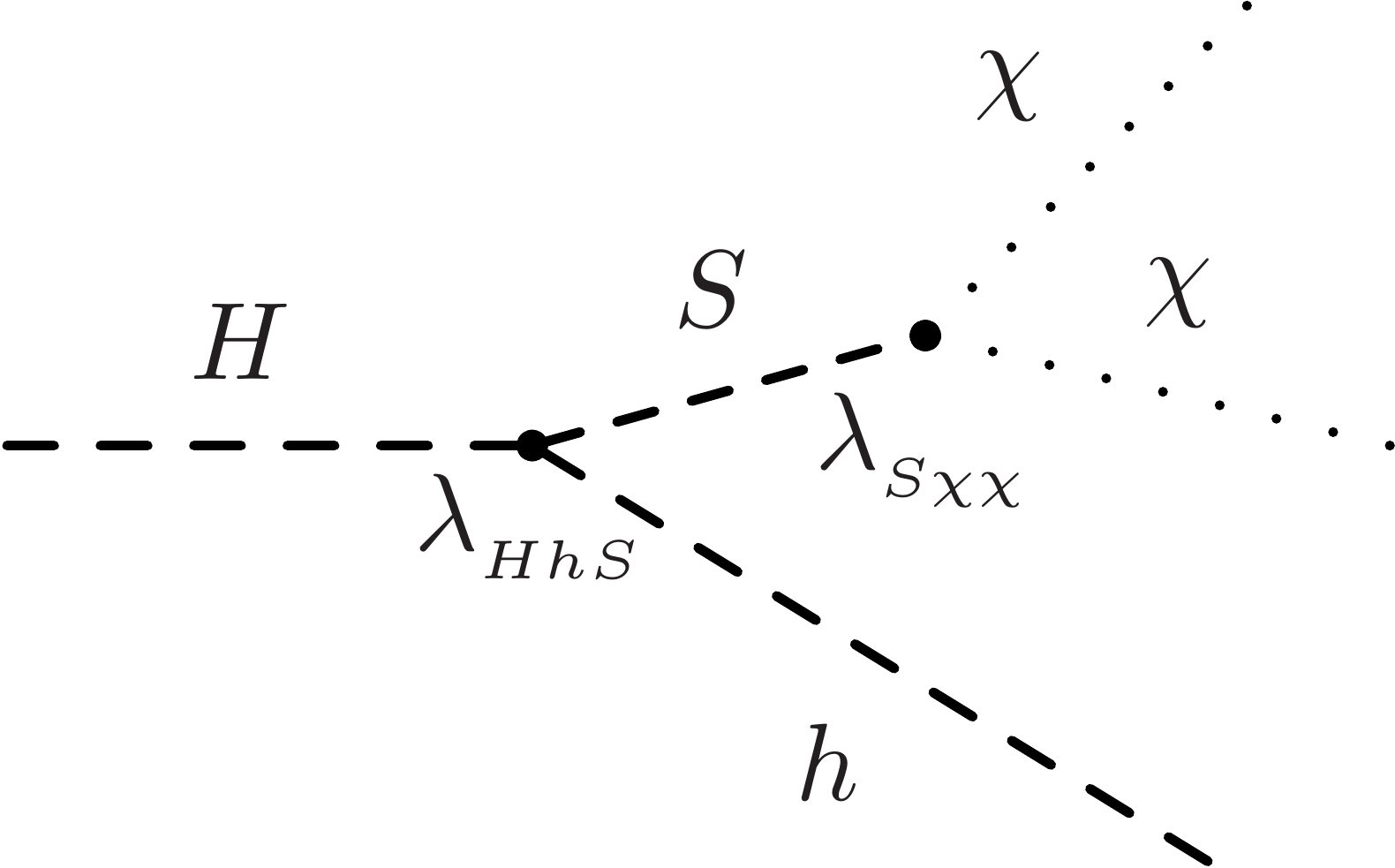}}
\caption{Higgs production in association with DM through the decay of $H$ using (a) an effective vertex and (b) the decay of a DM mediator $S$.}
\label{fig:diagrams}
\end{figure}

A first study on the hypothesis was done in 2015, where a new heavy scalar $H$ (the \textit{Madala\footnote{Madala is the Zulu word for an old man, the connotation being that $H$ is ``older'' and therefore heavier than the Higgs boson.} boson}) was introduced to explain several anomalous features in the LHC Run 1 data~\cite{vonBuddenbrock:2015ema}. In this past work, $H$ was considered to have Higgs-like couplings to the SM particles, as well as be a source of resonant di-Higgs production. It therefore was only considered in the mass range $2m_h<m_H<2m_t$, since anything heavier would have been dominated by $H\to t\bar{t}$ decays and anything lighter would not allow for resonant di-Higgs decays. This also lead to the assumption that $H$ is produced dominantly through gluon fusion ($gg$F), however the strength of the $g$-$g$-$H$ interaction was considered to be a free parameter, and the SM-like value could be rescaled by a factor $\beta_g^2$. The driving force behind the existence of $H$ was the Higgs \pt spectrum as measured by ATLAS in the $h\to\gamma\gamma$~\cite{Aad:2014lwa} and $h\to ZZ^*\to4\ell$~\cite{Aad:2014tca} decay channels. Both of these \pt spectra appeared to have a systematic enhancement in their intermediate bins (between 20 and 100~GeV). It was shown that this could be explained by requiring resonant associated Higgs production mode through the decay of $H$, i.e. $H\to h+X$. This was represented in terms of $h$ being produced in association with a pair of dark matter (DM) candidates, denoted by $\chi$, though an effective vertex represented in \autoref{fig:diagrams}(a). The DM candidate was a scalar, for simplicity, and fit all cosmological and detector constraints at a mass close to $\tfrac{1}{2}m_h$. With this model in place, a simultaneous fit was done to a collection of relevant ATLAS and CMS results, and a best fit mass of $H$ was found at $m_H=272^{+12}_{-9}$~GeV, with $\beta_g=1.5\pm0.6$. This result was obtained as a $3\sigma$ improvement of the SM prediction.

\begin{figure}
\centering
\includegraphics[width=350pt]{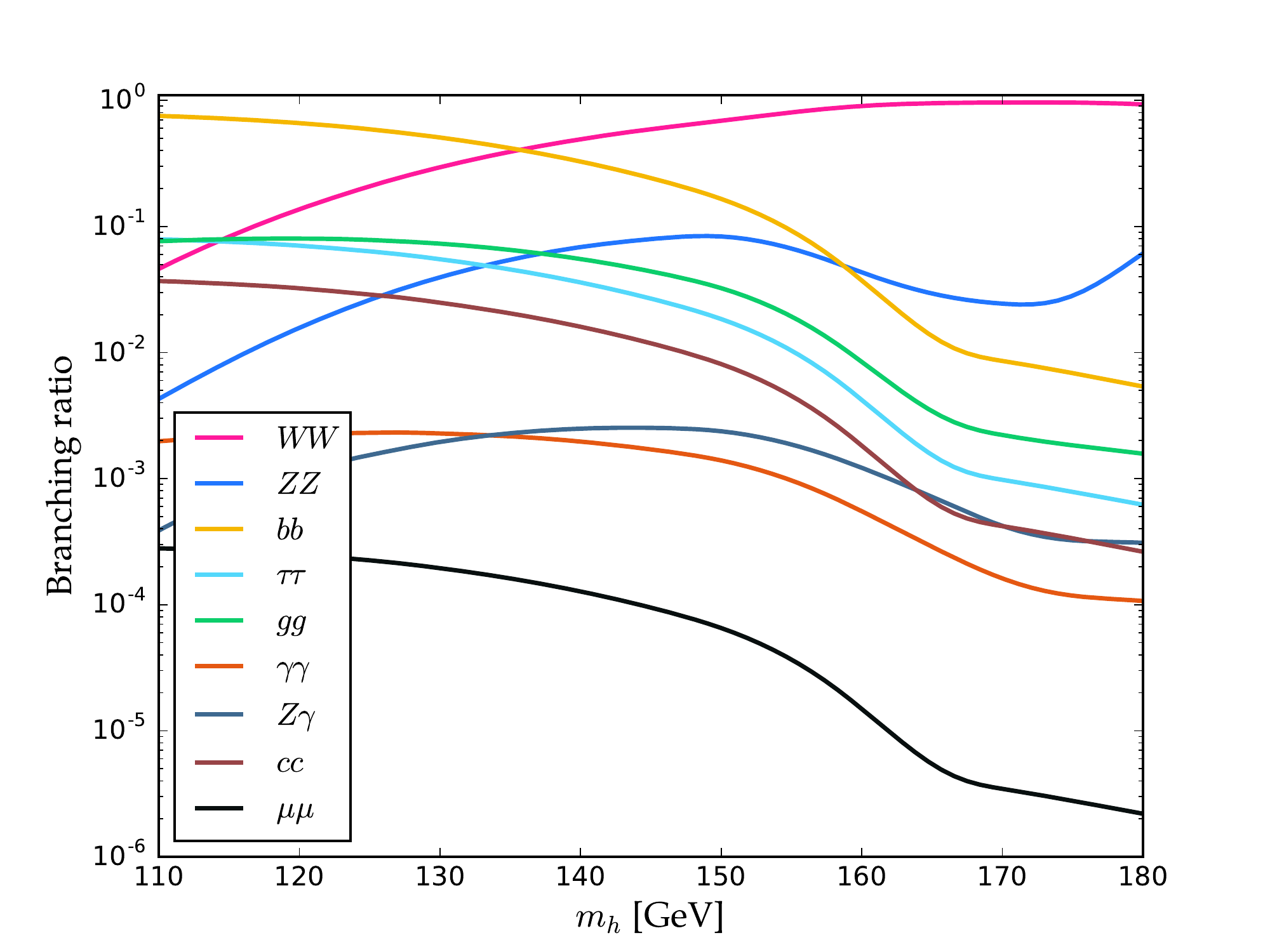}
\caption{The BRs of a Higgs-like boson in the mass range above and below $m_h$, taken from the LHC Higgs cross section working group~\cite{deFlorian:2016spz}. The $S$ boson (having a mass higher than $m_h$) would decay dominantly to the massive vector bosons increasingly depending on its mass.}
\label{fig:branchingratios}
\end{figure}

In 2016, a study was done~\cite{vonBuddenbrock:2016rmr} on determining how to understand the effective vertex in \autoref{fig:diagrams}(a). The results from the 2015 study seemed to indicate that the branching ratio (BR) of the $H\to h\chi\chi$ decay mode should have to be quite large to explain the data. It is not natural for a 3-body decay to have such a large BR. For this reason, the diagram shown in \autoref{fig:diagrams}(b) was proposed to explain the nature of the effective vertex. In this case, a scalar DM mediator $S$ was introduced, and the decays of $H$ changed such that the dominant modes are $H\to SS,Sh,hh$. The $S$ boson has a mass in the range $m_h<m_S<m_H-m_h$ such that it is more kinematically accessible through the decays of $H$ as mentioned above. It was also considered that $S$ has Higgs-like couplings to the SM, although its direct production is suppressed.\footnote{This assumption is merely a simplification to reduce the number of free parameters in the theory, and does not have a big impact on the BSM Higgs $p_\text{T}$ spectrum. It would, however, have a big impact on applying the Madala hypothesis to multilepton search results.} In this case, an $S$ boson with a mass around 160~GeV would decay dominantly to $W$ bosons, as can be seen in \autoref{fig:branchingratios}. The BRs to SM particles would, however, have to be suppressed by the $S\to\chi\chi$ BR, which is a free parameter of the theory.

The Madala hypothesis differs from many BSM hypotheses which predict heavy scalars, in that the Madala boson should dominantly decay to pairs of $h$ and $S$. The initial statistical study done in 2015 provided some insight into the potential parameter space of the model. However, since then a plethora of newer results from ATLAS and CMS have been presented (several of these at $\sqrt{s}=13$~TeV). For this reason, it is important to identify whether or not the results which are available at the time of writing this short paper are compatible with the results from the 2015 study.

\section{Statistical methodology\label{sec:stats}}

The experimental results which are relevant to study when considering the Madala hypothesis are shown in \autoref{tbl:searches}. It is apparent that such a diverse set of data and final states needs to be carefully combined in order for interesting information to be extracted. For this reason, a generic approach has been adopted in order to deal with statistics. That is, all experimental results are interpreted in terms of units of $\chi^2$. This simple approach is necessary due to the fact that not enough information is presented as part of experimental results.

The experimental results considered here are usually presented in two ways. Firstly, in the case where measurements are considered, a $\chi^2$ is calculated as Pearson's test statistic:
\begin{equation}
\chi^2=\frac{(\mu^\text{th}-\mu^\text{exp})^2}{(\Delta\mu^\text{th})^2 + (\Delta\mu^\text{exp})^2}.
\label{eqn:chisquare_measurement}
\end{equation}
Here, a theoretical prediction $\mu^\text{th}$ is compared against an experimental measurement $\mu^\text{exp}$, along with their respective uncertainties $\Delta\mu^\text{th}$ and $\Delta\mu^\text{exp}$. In the denominator, the experimental and theoretical uncertainties have already been added in quadrature since they are independent of each other.

Secondly, experimental results can come in the form of limits. For searches where no significant excess is seen, a 95\% CL is commonly what is presented. In this case, Pearson's test statistic is modified. The difference between the expected and observed limits are treated as a signal with a large error, and therefore limits contribute very weakly to a $\chi^2$. The contribution is written as follows:
\begin{equation}
\chi^2=\frac{(L^\text{obs}-L^\text{exp}-\mu^\text{th})^2}{(L^\text{exp}/1.96)^2},
\label{eqn:chisquare_limit}
\end{equation}
where $L^\text{exp}$ and $L^\text{obs}$ are the experimentally calculated expected and observed limits, respectively. Here again, $\mu^\text{th}$ is a theoretical prediction and its error is considered to be negligible compared with the experimental uncertainty, which is calculated as $L^\text{exp}/1.96$. The factor of 1.96 arises due to the fact that 95\% CL corresponds to 1.96 units of standard deviation. This method of calculating Pearson's test statistic has been tested in various cases, and found to be consistent with the standard definition given in \autoref{eqn:chisquare_measurement}, assuming that the calculated limits are statistically Gaussian.

Combinations of results can be performed using this $\chi^2$ method, and the results of the 2015 fit to data~\cite{vonBuddenbrock:2015ema} that constrained the parameters of the Madala hypothesis used this procedure. With this in mind, the newer search results can undergo the same treatment in order to understand whether or not the results are compatible with the 2015 fit result.

\begin{table}
\renewcommand{\arraystretch}{1.2}
\centering
\begin{tabular}{|c|c|c|c|}
	\hline
		\textbf{Result type} & \textbf{Collaboration} & \textbf{Run} & \textbf{Final state} \\
		\hline
		Higgs $p_\text{T}$  & ATLAS/CMS & 1 & $\gamma\gamma$, $ZZ^*\to4\ell$, $WW^*\to e\nu\mu\nu$~\cite{Aad:2014lwa,Khachatryan:2015rxa,Aad:2014tca,Khachatryan:2015yvw,Aad:2016lvc,Khachatryan:2016vnn} \\
		\cline{2-4}
		spectrum & ATLAS & 2 & $\gamma\gamma$~\cite{ATLAS:2016nke} \\
		 \hline
		 Di-Higgs & ATLAS & 1 & $bb\tau\tau$, $\gamma\gamma WW^*$, $\gamma\gamma bb$, $bbbb$~\cite{Aad:2015xja} \\
		 \cline{2-4}
		 & CMS & 1 & $bb\tau\tau$, $\gamma\gamma bb$, multilepton~\cite{Khachatryan:2015tha,CMS:2014ipa,Khachatryan:2014jya} \\
		 \cline{2-4}
		 & ATLAS & 2 & $\gamma\gamma bb$, $bbbb$, $\gamma\gamma WW$~\cite{ATLAS:2016ixk,ATLAS:2016qmt,TheATLAScollaboration:2016ibb} \\
		 \cline{2-4}
		 & CMS & 2 & $bb\tau\tau$,$\gamma\gamma bb$, $bbbb$, $bbWW$~\cite{CMS:2016knm,CMS:2016vpz,CMS:2016tlj,CMS:2016rec} \\
		 \hline
		 Di-boson & ATLAS/CMS & 1 & $WW$, $ZZ$~\cite{Khachatryan:2015cwa,Aad:2015kna,Aad:2015agg} \\
		 \cline{2-4}
		  & ATLAS & 2 & $ZZ\to4\ell,2\ell2\nu$, $WW\to e\nu\mu\nu$~\cite{ATLAS:2016kjy,ATLAS:2016oum,ATLAS:2016bza} \\
		  \cline{2-4}
		  & CMS & 2 & $ZZ\to2\ell2\nu$, $WW\to 2\ell2\nu$~\cite{CMS:2016jpd,CMS:2016noo} \\
		 \hline
	\end{tabular}
	\caption{A list of results that are relevant for consideration in the Madala hypothesis, up until before Moriond 2017. Several other final states are relevant too, such as top associated Higgs production results, but they are the focus of a future work. The 2015 fit result used all data from Run 1, with the exception of the $h\to WW^*\to e\nu\mu\nu$ $p_\text{T}$ spectra (for more details on the 2015 fit result, see Reference \cite{vonBuddenbrock:2015ema}).}
	\label{tbl:searches}
\end{table}

\section{Compatibility checks\label{sec:results}}

The fit result of the Madala boson mass in 2015 was found to be $m_H=272^{+12}_{-9}$~GeV, with $\beta_g=1.5\pm0.6$, as mentioned in \autoref{sec:intro}. Using this as a benchmark, one can use the statistical methodology described in \autoref{sec:stats} to combine the di-Higgs and di-boson search results in \autoref{tbl:searches} to try and understand whether the 2015 fit result is compatible with an updated dataset. These results all contain an interpretation that a heavy resonance $H$ is decaying to a pair of Higgs bosons (in the di-Higgs case) or a pair of massive vector bosons ($WW$ or $ZZ$, in the di-boson case). Since none of these results consider resonance masses lower that 260~GeV, the full parameter space considered for the Madala boson cannot be explored.

\begin{figure}
\vspace{-70pt}

\subfloat[]{\includegraphics[page=3,width=0.5\textwidth]{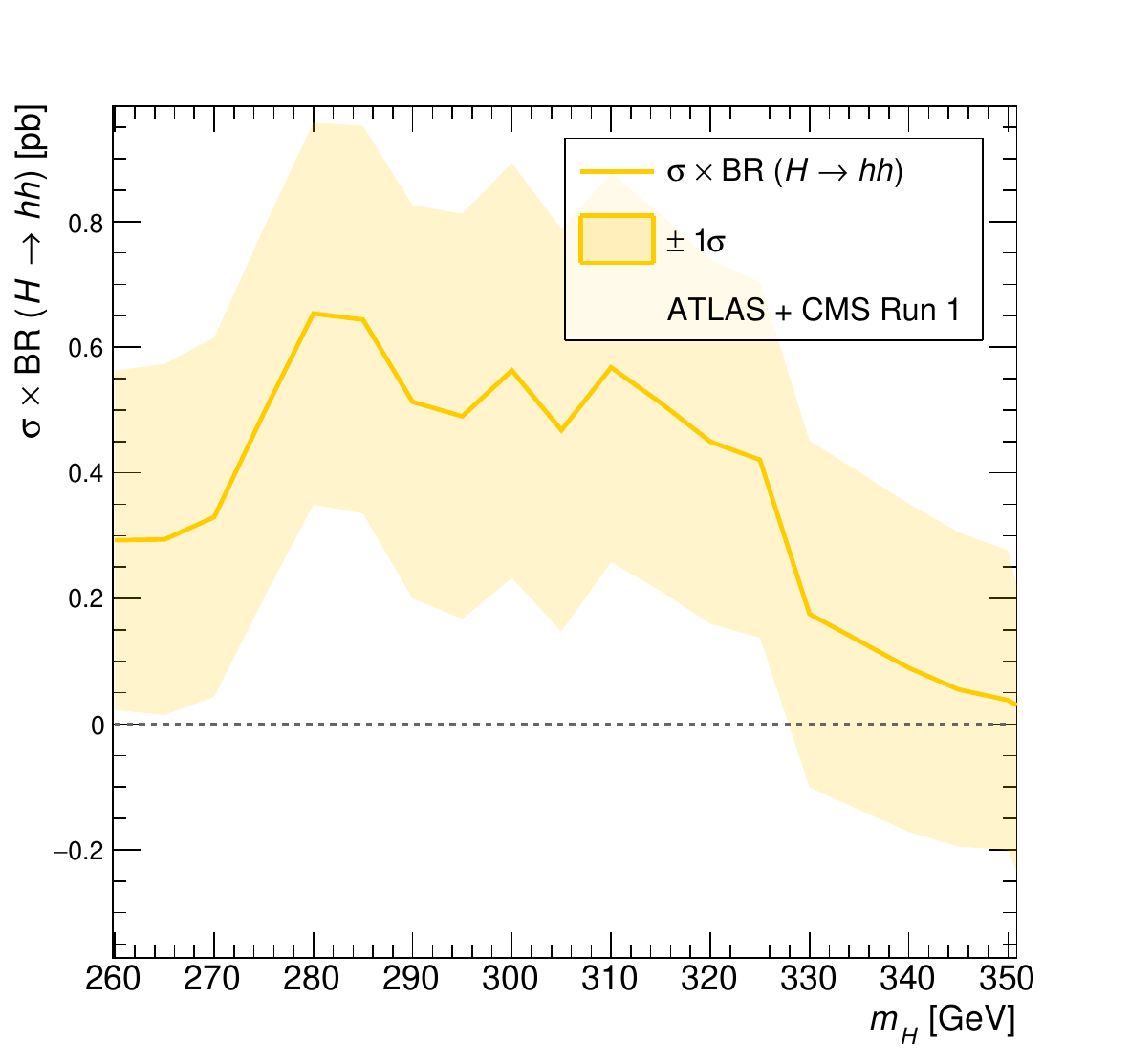}}
~
\subfloat[]{\includegraphics[page=6,width=0.5\textwidth]{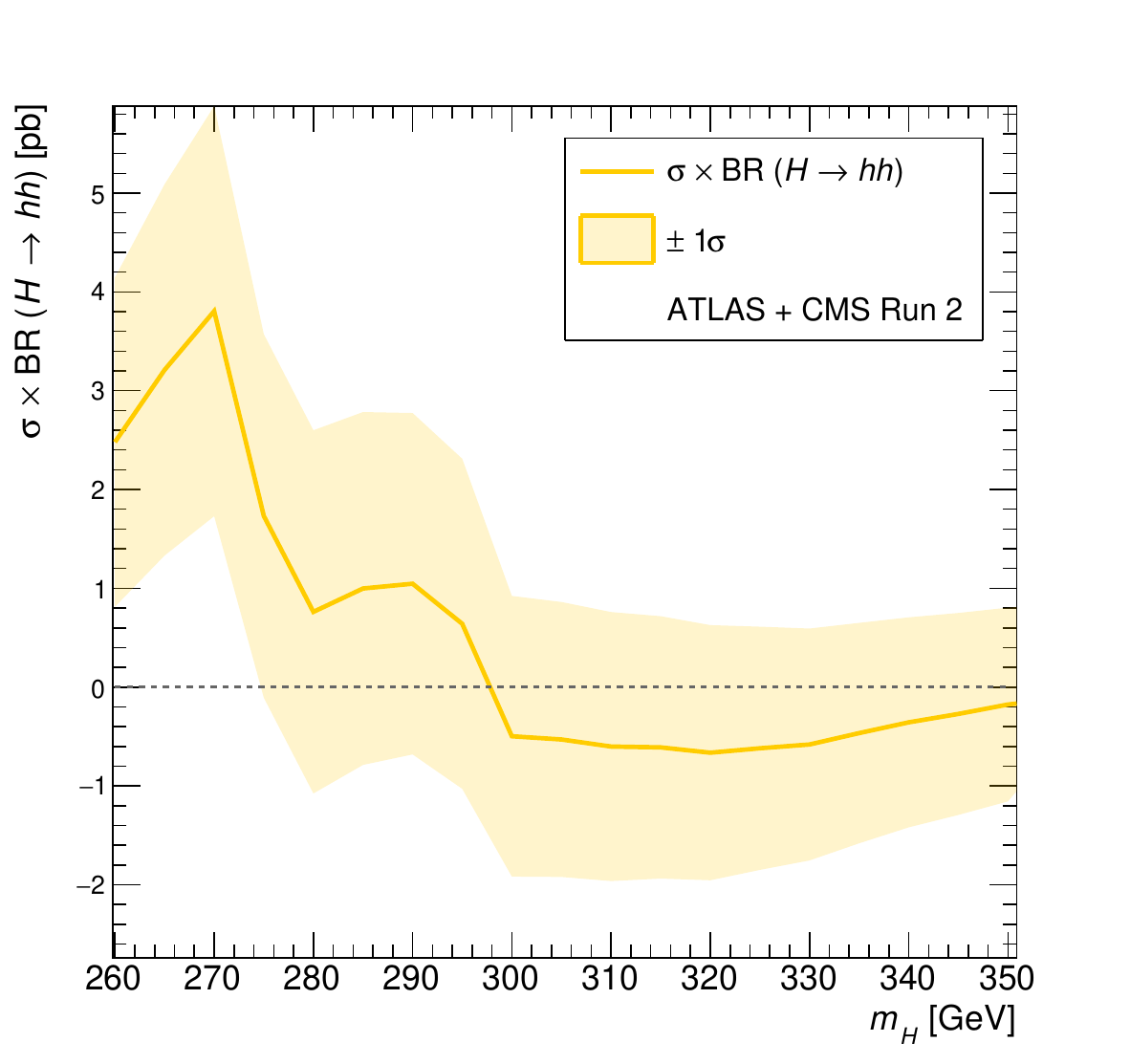}}
\vspace{-13pt}

\subfloat[]{\includegraphics[page=9,width=0.5\textwidth]{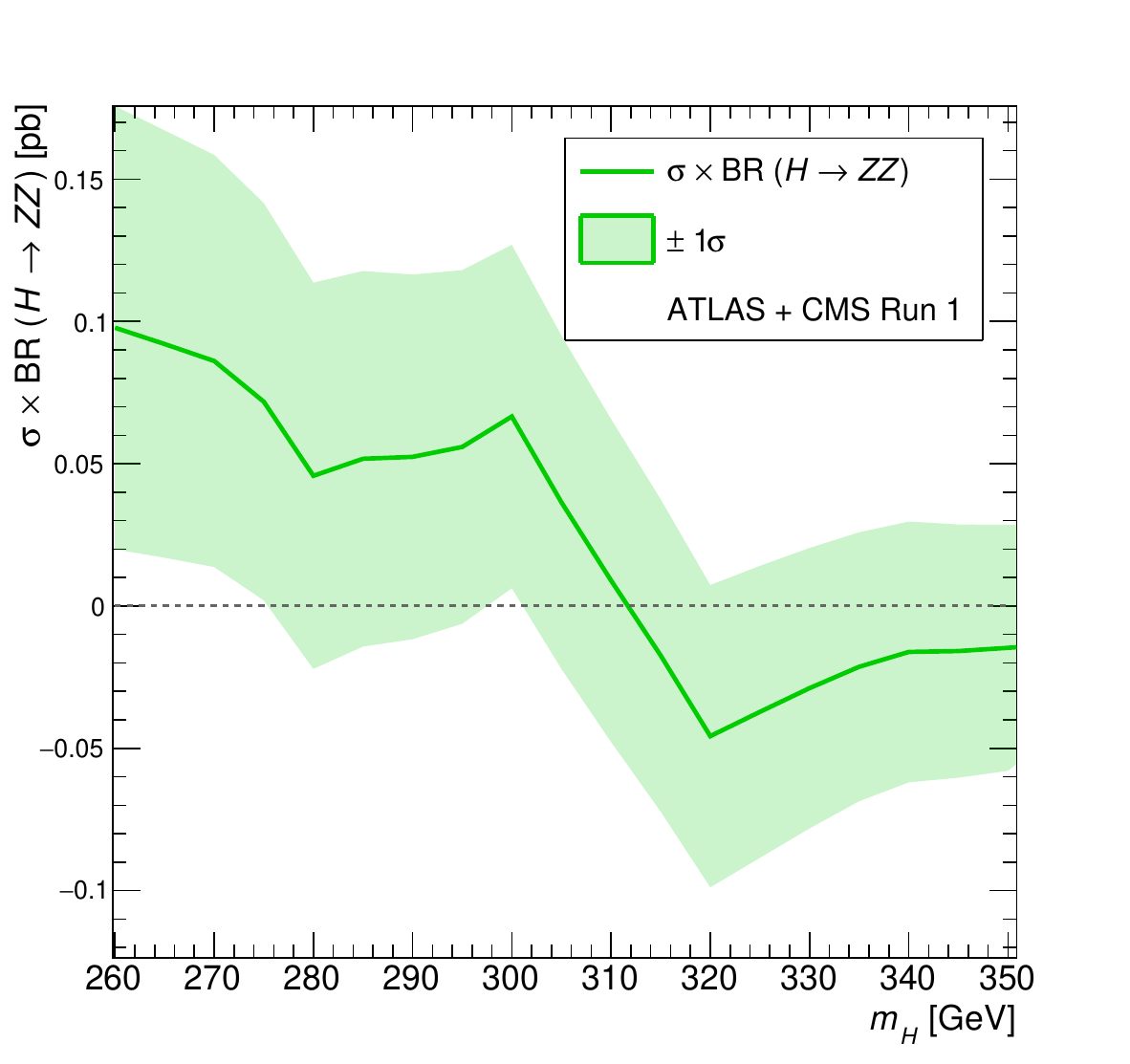}}
~
\subfloat[]{\includegraphics[page=12,width=0.5\textwidth]{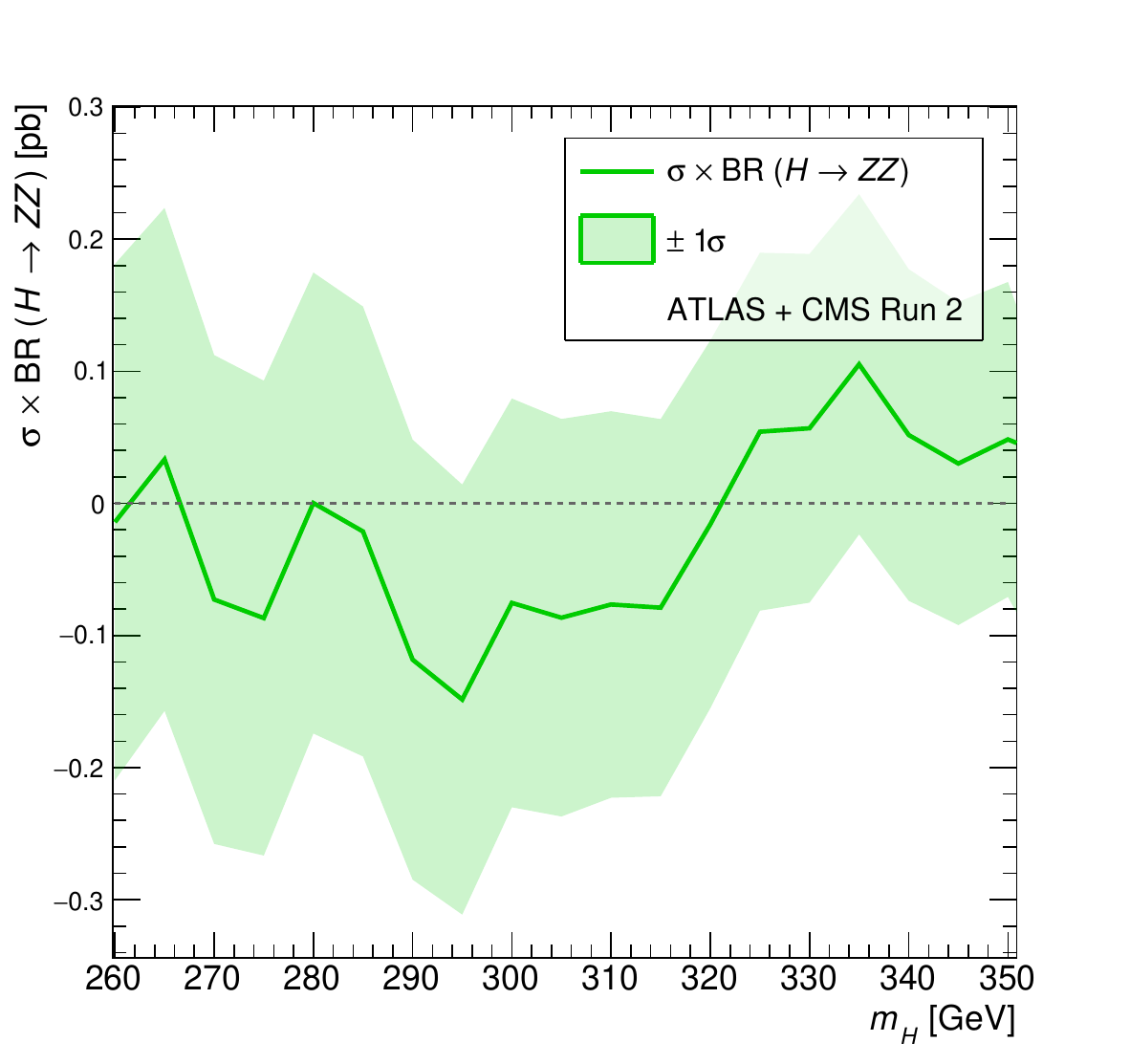}}
\vspace{-13pt}

\subfloat[]{\includegraphics[page=15,width=0.5\textwidth]{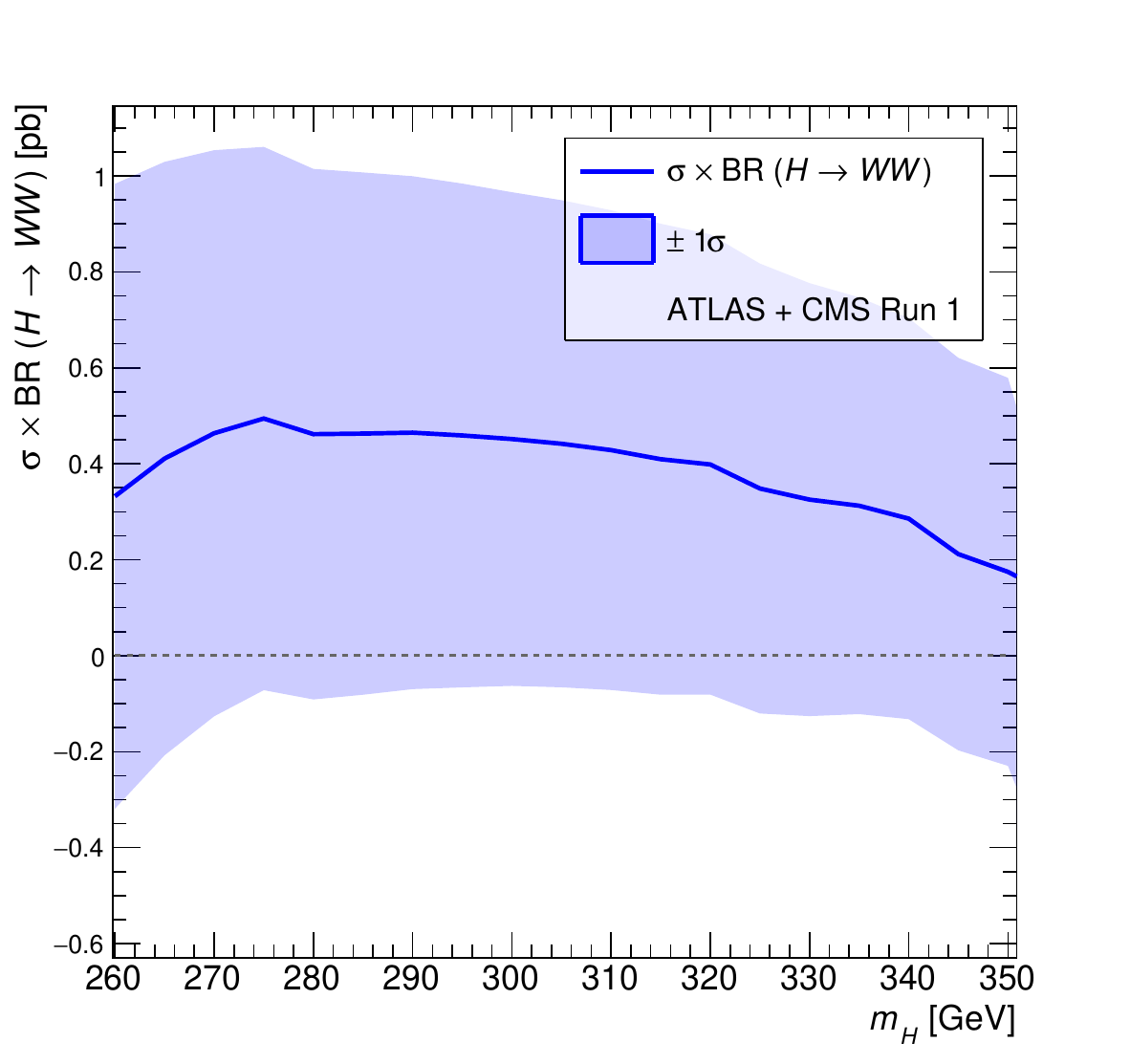}}
~
\subfloat[]{\includegraphics[page=18,width=0.5\textwidth]{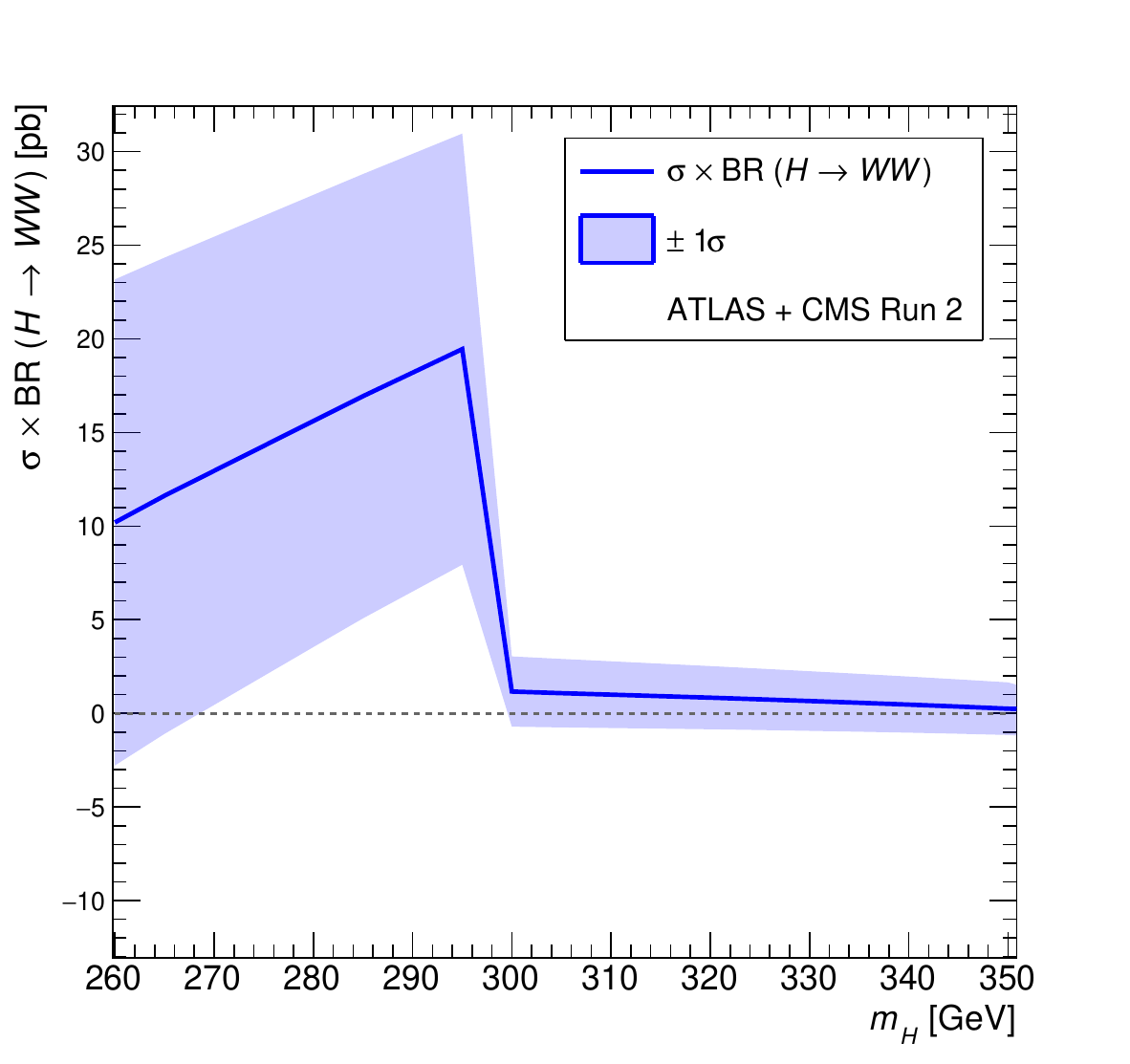}}
\caption{The best fit values of cross section times BR for $H$ production and decay into di-Higgs (top), $ZZ$ (middle), and $WW$ (bottom). The values have been separated into the 8~TeV Run 1 results (left) and the 13~TeV Run 2 results (right).}
\label{fig:searches}
\end{figure}

The results of a combination of the experimental data can be seen in \autoref{fig:searches}. On the vertical axis of each of these plots is a best fit value of cross section times BR for the associated search channel. On the horizontal axis the mass of the Madala boson is scanned. Bands have been drawn around the combined result, which represent a $1\sigma$ uncertainty in the result. Since we would expect differences in cross section for different center of mass energies, the results are separated into whether they come from Run 1 or Run 2. As can be seen, the combined result often deviates from the null hypothesis (i.e. that no resonance exists), and this most often happens in regions below $m_H=300$~GeV. The region around $m_H=272$~GeV shows an enhancement in cross section times BR in every case excepting the Run 2 $H\to ZZ$ result shown in \autoref{fig:searches}(d). By and large, these results are compatible with the 2015 fit result, and a more detailed study could provide us with a better constraint on the best fit value for $m_H$.

As mentioned above in \autoref{sec:intro}, the initial driving force behind the investigation of the Madala boson is the Higgs \pt spectrum. It is therefore also important to determine whether or not the 2015 fit result obtained with the Run 1 ATLAS and CMS $h\to\gamma\gamma$ and $ZZ^*\to4ell$ is compatible with what is obtained for the more recent results: the Run 1 ATLAS and CMS $h\to WW^*\to e\nu\mu\nu$ results and the Run 2 ATLAS $h\to\gamma\gamma$ result.

To study the Higgs \pt, a set of Monte Carlo (MC) samples were made to reproduce the different components of it. The SM Higgs \pt spectrum was separated into its different production mechanisms. The $gg$F spectrum was generated using the NNLOPS procedure~\cite{Hamilton:2015nsa}, which is accurate to next-to-next-to leading order (NNLO) in QCD. The associated production modes -- vector boson fusion (VBF), $Vh$ and $tth$ labelled together as $Xh$ -- were generated at next to leading order (NLO) using \textsc{MG5\_aMC@NLO}~\cite{Alwall:2014hca}. These spectra are scaled to the cross sections provided by the LHC Higgs Cross Section Working Group~\cite{deFlorian:2016spz} (from which the theoretical uncertainty also comes). The events are passed through an event selection identical to the fiducial selection used by the experimental collaborations. The dominant $gg$F prediction was further scaled by the experimentally measured signal strength for each decay mode, $\mu_{gg\text{F}}$.

\begin{table}[t]
\renewcommand{\arraystretch}{1.2}
\centering
	\begin{tabular}{|c|c|}
	\hline 
	Channel & $\beta_g^2$ \\
	\hline
	ATLAS Run 1 $h\to WW$ & $1.9\pm1.6$ \\
	ATLAS Run 2 $h\to\gamma\gamma$ & $1.0\pm1.7$ \\
	CMS Run 1 $h\to WW$ & $0$ \\
	\hline
	\end{tabular}
	\caption{Fit results for the study done on the Higgs \pt spectrum. Here the effective vertex shown in \autoref{fig:diagrams}(a) was used, with $m_H=270$~GeV and $m_\chi=60$~GeV.}
	\label{tbl:results}
\end{table}

\begin{figure}[b]
\subfloat[]{\includegraphics[page=10,width=0.5\textwidth]{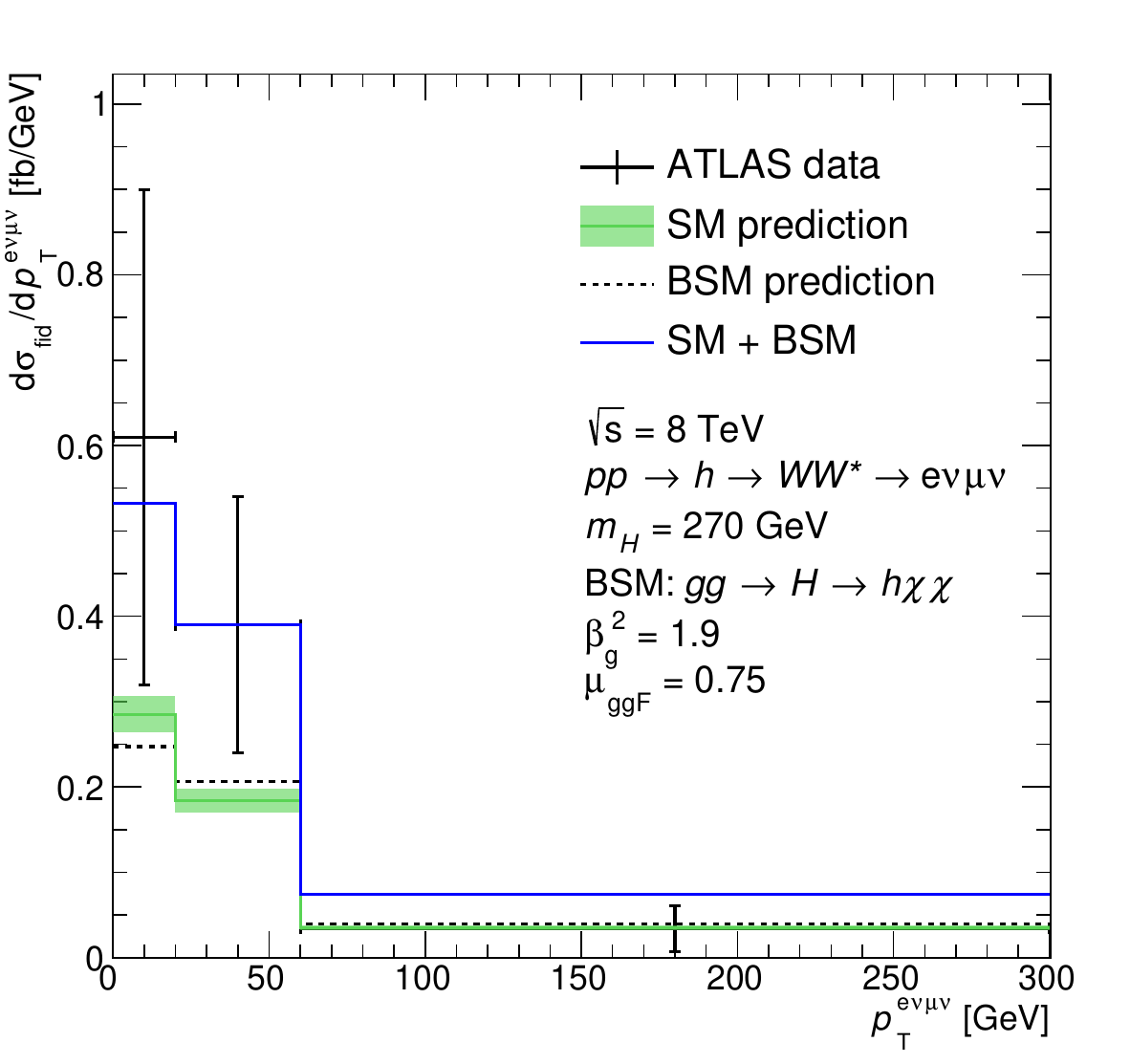}}
~
\subfloat[]{\includegraphics[page=11,width=0.5\textwidth]{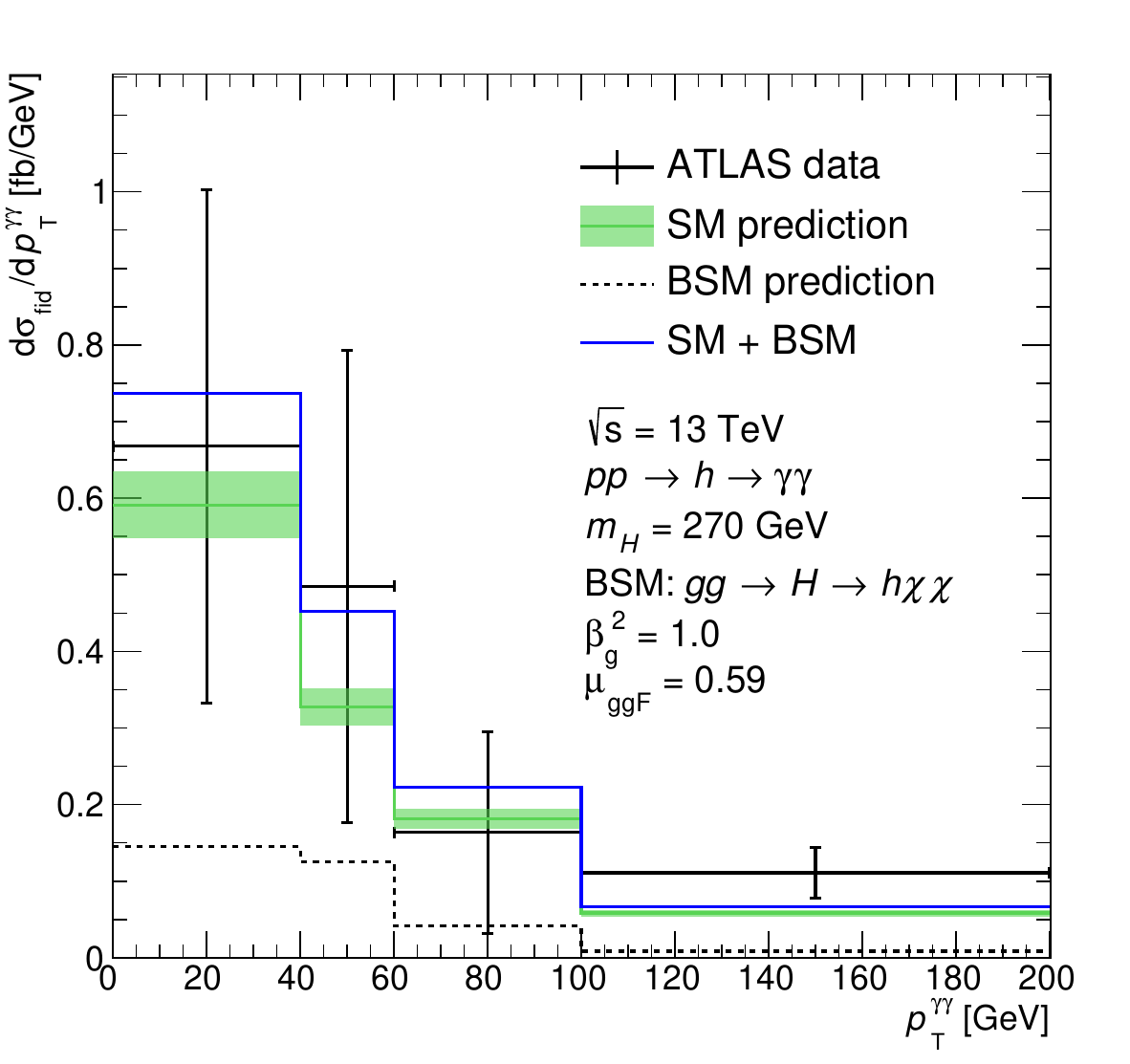}}
\caption{The Higgs \pt spectra for (a) the ATLAS Run 1 $h\to WW$ channel and (b) the ATLAS Run 2 $h\to\gamma\gamma$ channel. The mass points considered here are $m_H=270$~GeV and $m_\chi=60$~GeV, as described in the text.}
\label{fig:higgspt}
\end{figure}

The BSM prediction was considered to be the Madala hypothesis prediction of $gg\to H\to h\chi\chi$ through an effective vertex, as shown in \autoref{fig:diagrams}(a). This was generated using \textsc{Pythia 8.2}~\cite{Sjostrand:2014zea}, and scaled to the LHC Higgs Cross Section Working Group N$^3$LO $gg$F cross sections for a high mass Higgs-like scalar. The events were passed through the fiducial selections as in the case of the SM prediction. Since the Run 1 fit result had a best fit mass of $m_H=272$~GeV with $m_\chi=60$~GeV, the mass points considered for this study were $m_H=270$~GeV and $m_\chi=60$~GeV.

The SM and BSM components were added together and then tested for compatibility with the data. A $\chi^2$ value was calculated for each bin per channel, as in \autoref{eqn:chisquare_measurement}. The normalisation of the BSM spectrum is used to minimise the $\chi^2$, with the interpretation that it is scaled by the free parameter $\beta_g^2$. The results of this fit are shown in \autoref{tbl:results}. Comparing against the 2015 best fit point of $\beta_g^2=2.25$, the ATLAS Run 1 $h\to WW$ and ATLAS Run 2 $h\to\gamma\gamma$ results are compatible with this value. The CMS Run 1 $h\to WW$ is not improved by the BSM hypothesis. For reference, the \pt spectra at their best fit values are shown in \autoref{fig:higgspt} for the two spectra which are improved by the BSM hypothesis.

\section{The future of the Madala hypothesis\label{sec:conclusions}}

\begin{table}[t]
\vspace{-55pt}
\renewcommand{\arraystretch}{1.23}
\centering
\begin{tabular}{|c|c|c|}
\hline
Reference & Channel & Measured $\mu_{tth}$ \\
\hline
\multirow{4}{*}{CMS Run 1~\cite{Khachatryan:2014qaa}} & Same-sign $2\ell$ & $5.3^{+2.1}_{-1.8}$ \\
 & $3\ell$ & $3.1^{+2.4}_{-2.0}$ \\
 & $4\ell$ & $-4.7^{+5.0}_{-1.3}$ \\
\cline{2-3}
 & \textbf{Combination} & $2.8^{+1.0}_{-0.9}$ \\
\hline
\multirow{6}{*}{ATLAS Run 1~\cite{Aad:2015iha}} & $2\ell0\tau_\text{had}$ & $2.8^{+2.1}_{-1.9}$ \\
 & $3\ell$ & $2.8^{+2.2}_{-1.8}$ \\
 & $2\ell1\tau_\text{had}$ & $-0.9^{+3.1}_{-2.0}$ \\
 & $4\ell$ & $1.8^{+6.9}_{-2.0}$ \\
 & $1\ell2\tau_\text{had}$ & $-9.6^{+9.6}_{-9.7}$ \\
 \cline{2-3}
 & \textbf{Combination} & $2.1^{+1.4}_{-1.2}$ \\
\hline
\multirow{3}{*}{CMS Run 2~\cite{CMS:2017vru}} & Same-sign $2\ell$ & $1.7^{+0.6}_{-0.5}$ \\
 & $3\ell$ & $1.0^{+0.8}_{-0.7}$ \\
 & $4\ell$ & $0.9^{+2.3}_{-1.6}$ \\
\cline{2-3}
 & \textbf{Combination} & $1.5^{+0.5}_{-0.5}$ \\
\hline
\multirow{5}{*}{ATLAS Run 2~\cite{ATLAS:2016ldo}} & $2\ell0\tau_\text{had}$ & $4.0^{+2.1}_{-1.7}$ \\
 & $3\ell$ & $0.5^{+1.7}_{-1.6}$ \\
 & $2\ell1\tau_\text{had}$ & $6.2^{+3.6}_{-2.7}$ \\
 & $4\ell$ & $<2.2$ \\
\cline{2-3}
 & \textbf{Combination} & $2.5^{+1.3}_{-1.1}$ \\
\hline
\multicolumn{2}{|c|}{\textbf{Error weighted mean}} & $1.92\pm0.38$ \\
\hline
\end{tabular}
\caption{The measured $\mu$ values for $tth$ production in multileptonic analysis channels. A combination is estimated as the error weighted mean of each quoted combined result.}
\label{tbl:tth}
\end{table}

The Madala hypothesis is dependent on the availability of experimental results against which it can be tested. In 2015 a limited set of results was used, and its parameters were constrained using what was available at the time. The ATLAS and CMS collaborations, however, are always hard at work producing updated results with larger datasets. This short paper has compiled the results which were available before Moriond 2017, and shown that the newer results are compatible with the fit result obtained in 2015.

The results presented in this short paper are, however, not a complete re-fit of all available data. This task will be the focus of a future work, at a time where results are presented with the full 2015 and 2016 datasets. In addition to this, several aspects of the hypothesis have not been explored in this short paper. Most notably, the Madala hypothesis predicts that an enhanced rate of top associated Higgs production should be observed, particularly in leptonic channles, due to the existence of the $S$ boson. This could explain the enhancements we see in the data, the results of which are listed in \autoref{tbl:tth}.

As the LHC continues to run and larger datasets are analysed in the search for new physics, the Madala hypothesis will continue to be tested until such a point that we can confidently confirm it or rule it out. With several important results still to come out, more definite statements about the Madala hypothesis are left for a future work. 

\section*{References}

\bibliographystyle{iopart-num}
\bibliography{ref}

\providecommand{\newblock}{}
\begin{thebibliography}{10}
\expandafter\ifx\csname url\endcsname\relax
  \def\url#1{{\tt #1}}\fi
\expandafter\ifx\csname urlprefix\endcsname\relax\def\urlprefix{URL }\fi
\providecommand{\eprint}[2][]{\url{#2}}

\bibitem{Aad:2012tfa}
Aad G {\em et~al.\/} (ATLAS Collaboration) 2012 {\em Phys. Lett.\/} {\bf B716}
  1--29 (\texttt{arXiv}: \eprint{1207.7214})

\bibitem{Chatrchyan:2012xdj}
Chatrchyan S {\em et~al.\/} (CMS Collaboration) 2012 {\em Phys. Lett.\/} {\bf
  B716} 30--61 (\texttt{arXiv}: \eprint{1207.7235})

\bibitem{vonBuddenbrock:2015ema}
von Buddenbrock S, Chakrabarty N, Cornell A~S, Kar D, Kumar M, Mandal T,
  Mellado B, Mukhopadhyaya B and Reed R~G 2015  HRI-RECAPP-2015-011,
  WITS-MITP-010 (\texttt{arXiv}: \eprint{1506.00612})

\bibitem{Aad:2014lwa}
Aad G {\em et~al.\/} (ATLAS Collaboration) 2014 {\em JHEP\/} {\bf 09} 112
  (\texttt{arXiv}: \eprint{1407.4222})

\bibitem{Aad:2014tca}
Aad G {\em et~al.\/} (ATLAS Collaboration) 2014 {\em Phys. Lett.\/} {\bf B738}
  234--253 (\texttt{arXiv}: \eprint{1408.3226})

\bibitem{vonBuddenbrock:2016rmr}
von Buddenbrock S, Chakrabarty N, Cornell A~S, Kar D, Kumar M, Mandal T,
  Mellado B, Mukhopadhyaya B, Reed R~G and Ruan X 2016 {\em Eur. Phys. J.\/}
  {\bf C76} 580 (\texttt{arXiv}: \eprint{1606.01674})

\bibitem{deFlorian:2016spz}
de~Florian D {\em et~al.\/} (LHC Higgs Cross Section Working Group
  Collaboration) 2016  FERMILAB-FN-1025-T (\texttt{arXiv}: \eprint{1610.07922})

\bibitem{Khachatryan:2015rxa}
Khachatryan V {\em et~al.\/} (CMS Collaboration) 2016 {\em Eur. Phys. J.\/}
  {\bf C76} 13 (\texttt{arXiv}: \eprint{1508.07819})

\bibitem{Khachatryan:2015yvw}
Khachatryan V {\em et~al.\/} (CMS Collaboration) 2016 {\em JHEP\/} {\bf 04} 005
  (\texttt{arXiv}: \eprint{1512.08377})

\bibitem{Aad:2016lvc}
Aad G {\em et~al.\/} (ATLAS Collaboration) 2016 {\em JHEP\/} {\bf 08} 104
  (\texttt{arXiv}: \eprint{1604.02997})

\bibitem{Khachatryan:2016vnn}
Khachatryan V {\em et~al.\/} (CMS Collaboration) 2016 {\em Submitted to:
  JHEP\/} CMS-HIG-15-010, CERN-EP-2016-125 (\texttt{arXiv}:
  \eprint{1606.01522})

\bibitem{ATLAS:2016nke}
Aad G {\em et~al.\/} (ATLAS Collaboration) 2016  ATLAS-CONF-2016-067

\bibitem{Aad:2015xja}
Aad G {\em et~al.\/} (ATLAS Collaboration) 2015 {\em Phys. Rev.\/} {\bf D92}
  092004 (\texttt{arXiv}: \eprint{1509.04670})

\bibitem{Khachatryan:2015tha}
Khachatryan V {\em et~al.\/} (CMS Collaboration) 2015  (\texttt{arXiv}:
  \eprint{1510.01181})

\bibitem{CMS:2014ipa}
{CMS Collaboration} 2014  CMS-PAS-HIG-13-032

\bibitem{Khachatryan:2014jya}
Khachatryan V {\em et~al.\/} (CMS Collaboration) 2014 {\em Phys. Rev.\/} {\bf
  D90} 112013 (\texttt{arXiv}: \eprint{1410.2751})

\bibitem{ATLAS:2016ixk}
Aad G {\em et~al.\/} (ATLAS Collaboration) 2016  ATLAS-CONF-2016-049

\bibitem{ATLAS:2016qmt}
Aad G {\em et~al.\/} (ATLAS Collaboration) 2016  ATLAS-CONF-2016-071

\bibitem{TheATLAScollaboration:2016ibb}
Aad G {\em et~al.\/} (ATLAS Collaboration) 2016  ATLAS-CONF-2016-004

\bibitem{CMS:2016knm}
Khachatryan V {\em et~al.\/} (CMS Collaboration) 2016  CMS-PAS-HIG-16-029

\bibitem{CMS:2016vpz}
Khachatryan V {\em et~al.\/} (CMS Collaboration) 2016  CMS-PAS-HIG-16-032

\bibitem{CMS:2016tlj}
Khachatryan V {\em et~al.\/} (CMS Collaboration) 2016  CMS-PAS-HIG-16-002

\bibitem{CMS:2016rec}
Khachatryan V {\em et~al.\/} (CMS Collaboration) 2016  CMS-PAS-HIG-16-011

\bibitem{Khachatryan:2015cwa}
Khachatryan V {\em et~al.\/} (CMS Collaboration) 2015 {\em JHEP\/} {\bf 10} 144
  (\texttt{arXiv}: \eprint{1504.00936})

\bibitem{Aad:2015kna}
Aad G {\em et~al.\/} (ATLAS Collaboration) 2016 {\em Eur. Phys. J.\/} {\bf C76}
  45 (\texttt{arXiv}: \eprint{1507.05930})

\bibitem{Aad:2015agg}
Aad G {\em et~al.\/} (ATLAS Collaboration) 2016 {\em JHEP\/} {\bf 01} 032
  (\texttt{arXiv}: \eprint{1509.00389})

\bibitem{ATLAS:2016kjy}
Aad G {\em et~al.\/} (ATLAS Collaboration) 2016  ATLAS-CONF-2016-074

\bibitem{ATLAS:2016oum}
Aad G {\em et~al.\/} (ATLAS Collaboration) 2016  ATLAS-CONF-2016-079

\bibitem{ATLAS:2016bza}
Aad G {\em et~al.\/} (ATLAS Collaboration) 2016  ATLAS-CONF-2016-056

\bibitem{CMS:2016jpd}
Khachatryan V {\em et~al.\/} (CMS Collaboration) 2016  CMS-PAS-HIG-16-023

\bibitem{CMS:2016noo}
Khachatryan V {\em et~al.\/} (CMS Collaboration) 2016  CMS-PAS-HIG-16-001

\bibitem{Hamilton:2015nsa}
Hamilton K, Nason P and Zanderighi G 2015 {\em JHEP\/} {\bf 05} 140
  (\texttt{arXiv}: \eprint{1501.04637})

\bibitem{Alwall:2014hca}
Alwall J, Frederix R, Frixione S, Hirschi V, Maltoni F, Mattelaer O, Shao H~S,
  Stelzer T, Torrielli P and Zaro M 2014 {\em JHEP\/} {\bf 07} 079
  (\texttt{arXiv}: \eprint{1405.0301})

\bibitem{Sjostrand:2014zea}
Sjostrand T, Ask S, Christiansen J~R, Corke R, Desai N, Ilten P, Mrenna S,
  Prestel S, Rasmussen C~O and Skands P~Z 2015 {\em Comput. Phys. Commun.\/}
  {\bf 191} 159--177 (\texttt{arXiv}: \eprint{1410.3012})

\bibitem{Khachatryan:2014qaa}
Khachatryan V {\em et~al.\/} (CMS Collaboration) 2014 {\em JHEP\/} {\bf 09} 087
  [Erratum: JHEP10,106(2014)] (\texttt{arXiv}: \eprint{1408.1682})

\bibitem{Aad:2015iha}
Aad G {\em et~al.\/} (ATLAS Collaboration) 2015 {\em Phys. Lett.\/} {\bf B749}
  519--541 (\texttt{arXiv}: \eprint{1506.05988})

\bibitem{CMS:2017vru}
Khachatryan V {\em et~al.\/} (CMS Collaboration) 2017  CMS-PAS-HIG-17-004

\bibitem{ATLAS:2016ldo}
Aad G {\em et~al.\/} (ATLAS Collaboration) 2016  ATLAS-CONF-2016-058

\end{thebibliography}
\end{document}